# Three Dimensional Subwavelength focus by a near-field plate lens

Lu Lan, Wei Jiang, and Yungui Ma*

*Centre for Optical and Electromagnetic Research, State Key Lab of Modern Optical Instrumentation, Zhejiang University, Hangzhou 310058, China*

**Abstract**
We implemented the inverse design method to build a thin near-field lens that could produce a desired subwavelength focus by manipulating the near fields of a magnetic dipole source. The flat near-field lens represented by an artificial impedance surface was fabricated by lumped reactive elements (capacitor and inductor) with spatially varying values. In the experiment a desired annular focusing spot with a characteristic size nearly three times smaller than that allowed by the diffraction limit was obtained. Besides high-resolution imaging, the proposed near-field plate could be extended for other interesting applications, such as wireless power transfer or complex wavefront/beam shaper.

*Corresponding e-mail: yungui@zju.edu.cn

**Introduction**

Recently thin artificial electromagnetic (EM) plates (or meta-surfaces) engineered with periodic subwavelength scatters have gained great research interests in the community since the first work reporting the generalized Snell's law [1-10]. In optics the patterned meta-surfaces are usually composed of subwavelength antenna arrays and configured to interact with the incident light by producing specific spatial phase/amplitude modulations on the scattered waves. Different from the traditional spatial light modulator (SLM) made of thick anisotropic liquid crystals, a compact and monolithic meta-surface device can modulate the spatial wavefront pattern in a deep subwavelength scale and manipulate the wave refraction/reflection behaviors without suffering the diffraction limit. This unique characteristic has already led to some interesting applications such as a transformer to link propagating to surface waves [11] or a shaper to produce a near-field vortex beam [12] which may excite some new technologies, as analyzed in a recent review article, in the fields such as energy harvesting or on-chip optical devices [13].

In this letter we report our experiment work on another type of meta-surface device that is able to focus a given source profile into a deep subwavelength spot. The design theory was first introduced by Merlin based on an inverse constructing scheme [14]. It works on the evanescent waves so that is called by a near-field flat lens in this paper. The overall structure, though similar to Fresnel plates which force the input field to converge to a spot in the focal plane [15] for the interference of traveling waves, performs with a quasi-static form of interference [14]. The first experimental device was demonstrated in two dimensions (2D) by Grbic et al showing a diffractionless near-field flat lens in microwaves [16,17]. They utilized spatially lumped electric elements (capacitors) to build the impedance strip lens, which have the equivalent functions as an optical antenna array. Later they designed a thick flat lens made of corrugated metallic surface and realized a subwavelength focus in three dimensions (3D) [18]. But their practical application is limited by a relative small aperture size due to the fact that the design theory originating from the coupled inductive rings is valid only at a quasi-static precondition [19]. In this work we overcome this problem by developing an artificial impedance surface lumped with spatially varying reactive elements for 3D subwavelength focus. Our near-field plate lens is constructed on the application of the strict solutions of Maxwell's boundary equations and in principal can be engineered without a limit on the aperture size. The surface design is compact and monolithic, which could be readily realized by a standard lithography technology. The measurement shows our lens has a prominent subwavelength focusing capability. Other technological applications such as enhancing the efficiency of wireless power transfer may be envisioned based on the current results [20].

**Design the near-field plate lens**

We firstly brief the inverse construction theory to design a near-field lens that is able to focus the near fields of a magnetic dipole. According to the theory, a desired focal field distribution is first defined and then expanded to retrieve the transmission field on the near-field plate lens through back propagation theory. With the incident and transmission fields known, the impedance profile of the near-field plate can then be calculated by a circuit equation derived from the boundary consistent condition [17]. The process is schematically shown in Fig. 1(a) where it shows a near-field plate lens at $z = 0$ can transform the field of a magnetic dipole (i.e., a subwavelength current loop) at $z = -L$ into a pre-defined subwavelength spot in the focal plane at $z = L$.

For the simplicity of implementation, we assume our lens of a rotational symmetry by applying a magnetic dipole source and thus only need to consider the tangential electric component $E_\phi$. In the cylindrical coordinate the harmonic EM field with a time factor $exp(i\omega t)$ ($\omega$ is the angular frequency) could be expressed in the angular-representation form $F(q_r, z_\alpha)$ by

$$F(q_r, z_\alpha) = \int_0^{+\infty} E(r, z_\alpha) J_1(q_r r) r dr, \qquad (1)$$

where $q_r$ is the wavevector along the radial direction and $J_1(x)$ is the first kind Bessel function of the first order. To acquire a subwavelength focusing profile, wave components of large wavevectors (i.e. $q_r \gg k_0$, where $k_0$ is the wavevector in free space) need to be taken into account. As shown in the inset of Fig. 1(b), the spatial spectrum of our predefined focus field is selected to have a uniform distribution from $q_1$ to $q_0$ which satisfies $k_0 \ll q_1 \ll q_0$. From Eq. (1) the spatial field distribution is calculated by applying the inverse Fourier transformation and the result is given in Fig.1 (b) as denoted by the blue circle line. The full width at the half maximum (FWHM) of the peak describing the focusing capability for an ideal case is about $0.12\lambda$ ($\lambda$ is the wavelength in free space), far smaller than the diffraction limit (at about $\lambda$ for a annulus profile).

Applying the back propagation theory, we obtain the transmission field of the near-field lens in the plane at $z = 0$ by the equation

$$E_\phi(r, z_\alpha) = \int_0^{+\infty} F(q_r, z_\beta) J_1(q_r r) \times e^{i[q_r(r-r') + \kappa(z_\alpha - z_\beta)]} q_r dq_r \qquad (2),$$

where $\kappa$ is the propagation constant along the $z$ direction which satisfies

$$\kappa = \begin{cases} \sqrt{k_0^2 - q_r^2} & (k_0 > q_r) \quad (3\text{-a}) \\ i\sqrt{q_r^2 - k_0^2} & (k_0 < q_r) \quad (3\text{-b}) \end{cases}.$$

With the incident ($E_i$) and transmission ($E_t$) fields available on the two sides of the plate lens, the next step is to calculate the impedance of this surface to realize the predefined transformation relationship from the boundary consistent condition. For a structure of rotational symmetry, the boundary condition can be written by

$$E_i(r) + \int_0^R J_1(r')C_{loop}(r',r)r'dr' = E_t(r), \qquad (4)$$

where $C_{loop}(r',r)$ represents the electric field at the radius $r$ contributed by a differential current loop of an average radius $r'$ in the polar coordinate. There is no near-field analytical solution for Eq. (4) due to the complexity of $C_{loop}(r',r)$ [21,22]. On the other hand we can obtain its numerical solution by resorting to a partial differential equation (PDE) solver (COMSOL). Consequently the distribution of current density $J_\phi(r)$ is solved by discretizing the integral on the left side of Eq. (4), which relates to the required surface impedance $Z_s(r)$ through a simple boundary circuit equation

$$E_t(r) = Z_s(r)J_\phi(r). \qquad (5)$$

In this work our main interest is to utilize this inverse design and patterned surface technology to manipulate evanescent waves and realize a deep subwavelength focus in the near field. We choose the frequency to be 3.0 GHz ($\lambda$ = 100 mm, equal to the radius of our implemented lens). In the experiment, as schematically shown in Fig. 1(a), a current loop (i.e., source) of radius = $0.05\lambda$ is placed at a distance $L$ to the flat lens, where $L$ (=$0.1\lambda$) is equal to the focal length $f$. Figure 2(a) gives a cartoon picture of our flat lens. The ideal lens model is approximated by 30 discrete homogeneous concentric rings which have equal widths (i.e., $\lambda/30$). Each ring is again approximated by $3N$ homogenous subwavelength segments along the tangential direction ($N$ is the ring order counted from the center). In the impedance calculation we use rectangular elements (width $w = \lambda/30 = 3.33mm$ and height $h = 2\pi/3 \times w = 6.97mm$) to approximate these curved segments, which is usually employed in the unit design of metamaterial devices [23, 24].

The calculated complex impedance values for each rings are listed in Table I as Fig. 2(b). In the experiment we approximate these complex numbers by taking their imaginary parts since most of them have negligible real parts and then implement them by reactive elements consisting of inter-digital capacitor or wire inductor printed on a thin circuit board (Rogers 6006), as shown in the inset of Fig. 2(a). Different impedance values are realized by adjusting the finger's length or spacing. For the rings of $N$ = 13, 16, 22 and 28 which have comparable real and imaginary impedance values, we approximate the original complex $Z_s$ by a pure imaginary number but keeping the sign and modulus unchanged. These treatments slightly modify the final focal profile by introducing small side lobes, as shown in Fig. 1(b) by the red triangle line. But the main feature of the subwavelength focus is not affected.

**Results and discussions**

Our near-field lens was fabricated by a standard lithography technique. The small current loop (radius = $0.05\lambda$) was made by a thin copper wire fed by a coaxial cable. It needs to mention that experimentally we are not able to make an ideal magnetic dipole but an approximated one with the existence of a tiny gap to feed the energy. This imperfection will induce some undesired electric components ($E_z$ and/or $E_r$) besides the desired tangential field ($E_\Phi$). In measurement the source and the lens were mounted on a step-motor-driven platform stage segregated by a 300 mm microwave absorber and the whole system was placed inside a microwave anechoic room. A monopole antenna was used to probe the near field signal, which was generated and processed by a vector network analyzer (RS-ZVA40). The step length for scanning the spatial field was controlled at 1 mm and the measurement frequency varies from 1.0 to 5.0 GHz.

From the measurement data, we found the best focus performance was achieved at 3.2 GHz, slightly deviated from the designed frequency of 3.0 GHz, which could be caused by the imperfect fabrication of our lens and the approximations used as well. Figures 3(a) to 3(d) show the analytical (left column) and the measured (right column) field profiles in the focal plane at $z = f = 0.1\lambda$ with (bottom row) and without (top row) the near-field plate lens. The measured field profile with the lens [Fig. 3(d)] is indeed substantially narrowed in comparison with the case without lens [Fig. 3(b)], indicating the focusing function of our lens. Compared to the analytic results, our measured fields also contain the contributions of the undesired components caused by the imperfect source as discussed earlier and thus show broader widths. This point is evidently seen from the near-field patterns for the empty case (i.e., no lens) as shown in Figs. 3(a) and 3(b) where the experimental field has a larger spatial distribution. Nonetheless subwavelength focus for the near fields is obtained by our near-field plate lens.

The qualities of our lens and the source is further examined from the electric field profiles plotted along a straight line passing the spot center in Figs. 3(a) to 3(d). The results are given in Figs. 4(a) and 4(b) for the cases without and with lens, respectively. As shown in Fig. 4(a), the measured source field (red triangle line) is in good agreement with that for the ideal source (blue dot line) except for the small region around the center as indicated by a non-zero value, which manifests the existence of other components ($E_z$ and/or $E_r$). With the lens placed on the source, as shown in Fig. 4(b), the line profiles are greatly narrowed. The FWHM value of the experimental data is estimated to be about $0.34\lambda$, which is roughly three times larger than the designed value ($0.12\lambda$) but nearly three times smaller than the case without lens ($0.97\lambda$). Thus our near-field plate lens has a compromised focus capability with a focusing index 2.85. This value is acceptable when consider the approximation on the impedance values we adopted and the imperfect source used. We believe the focusing capacity could be further enhanced by incorporating resistance elements in the fabrication [25].

We also examined the field patterns at other positions to have a deeper understanding on the spatial focus characteristic of our lens. Figures 5(a) to 5(c) showed the field pictures at $d = 7$, 10 and 13 mm, respectively, where $d$ is the distance to the lens surface from the measured plane. From these figures we can see that the field in the focal plane ($d = f = 10$ mm) has the narrowest distribution, which is more explicitly shown in Fig. 5(d) by plotting the field profile across the spot center. These results are consistent with our design. Figure 5(d) also shows there exists certain focusing depth for our lens around the focal plane, which is important for practical application. Here we would like to emphasize that the magnetic fields in the near-field planes are simultaneously concentrated from the principal of duality. This will be a very interesting result for the technology of wireless power transfer which mainly deals with the inductive coupling of magnetic field in the near field [20].

**Conclusions**

In this work, we have first built a thin near-field plate lens to realize the subwavelength focus of a three-dimensional source field enabled by the direct operation on the evanescent waves. Experimentally we achieved a resolution enhancement in the focal size by a factor of nearly 3. The quality of our device is determined by the condition that how accurate the impedance profile of the meta-plate could be implemented. This will impose certain constraints on the field patterns of the incident and transmission fields. But in principal complex impedance surface could be implemented by lumping both resistors and reactive elements. In optics the inverse design method may be also applied to generating some desired focus beams facilitated by the optical components such as nano inductor or nano capacitor [26, 27]. This may help to overcome the limits in aperture size or focal distance encountered by using the plasmonic devices such as centrally holed bullseye gratings in achieving a near-field subwavelength spot [28,29].

Lastly we would like to emphasize that our near-field plate design is based on the strict solution of the inversed boundary equations [i.e., Eqs (4) and (5)] and theoretically able to precisely transform the known incident field into the pre-defined output pattern by an impedance device of compactly lumped elements. The previous similar work on the meta-surface lens neglected the interaction among the subwavelength antennas [13] and had a finite resolution in modulating the spatial phase profile limited by the maximum density of the constituent antennas. These will not be issues in our design and fabrication. Besides for the subwavelength focus, the meta-plate impedance lens with the inverse design can be managed to produce diverse wave effects that may lead to some other important applications such as wireless power transfer or a wave-front/beam shaper.

**Acknowledgment**

This work is partially supported by the grants of NSFCs 61271085, 60990322 and 91130004, the National High Technology Research and Development Program (863 Program) of China (No. 2012AA030402), NSF of Zhejiang Province (LY12F05005), NCET, MOE SRFDP of China, Scientific Research Fund of Zhejiang Provincial Education Department.

**Captions**

FIG. 1 Diagram of our near-field lensing system with the desired focus profile. (a) The lensing system: a near-field plate lens at $z = 0$ will focus the EM fields of a current loop at $z = -L$ into a deep subwavelength spot in the focal plane at $z = L$. (b) The predefined normalized electric field profiles for the focus spot with the blue circle line representing the ideal analytic profile and the red triangle line representing a modified profile after taking the impedance approximation into account. The inset in (b) plots the wavevector spectrum of the target focus we defined.

FIG. 2 Design the near-field plate lens. (a) A cartoon picture of our designed near-field plate lens and the inset showing a zoom-in local picture for an inductor and a capacitor. (b) Table of the numerically obtained impedance values for the $N^{th}$ ring counted from the center.

FIG. 3 Electric field modulus patterns in the focal plane. (a) and (b) Correspond to the analytical and measured field profiles without near-field lens, respectively. (c) and (d) Correspond to the analytical and measured field profiles with near-field lens, respectively. The spatial fields are normalized by the peak value and so do for the figures 5.

FIG. 4 Cross-section line profile of the focus field. (a) The line profile is plotted along a straight line passing the spot center in Figs. 3(a) and 3(b). (b) The line profile is plotted from Figs. 3(c) and 3(d). The red triangle and the blue dot represent the theoretical and the experimental results, respectively.

FIG. 5 Electric field modulus patterns at different planes. (a) Field pattern at $d = 0.7f$. (b) Field pattern at $d = f$. (c) Field pattern at $d = 1.3f$. (d) Cross-section line profile plotted along a straight line crossing the spot center in (a), (b) and (c) as indicated by the line's legends

**FIGURE 1**

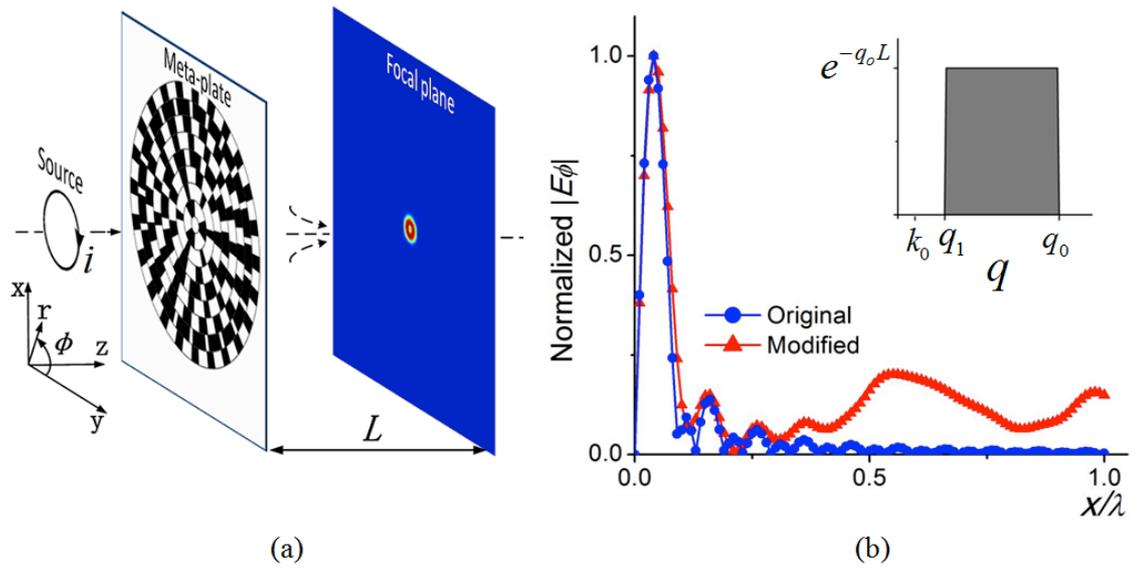

(a)                          (b)

FIGURE 2

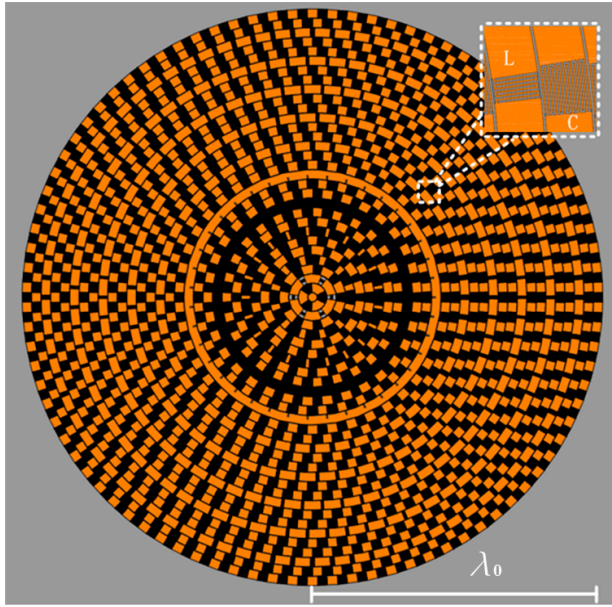

TABLE I. $Zs$ ($\Omega$) on $N^{th}$ ring counted from center

| No. | Zs | No. | Zs |
|---|---|---|---|
| 1 | 0.048-17.4 I | 16 | 22.1+5.91 I |
| 2 | 0.183+13.8 I | 17 | -0.613-15.4 I |
| 3 | -0.076-17.7 I | 18 | 0.619-14.4 I |
| 4 | 0.073-10.9 I | 19 | 30.5-12.3 I |
| 5 | 0.192-21.0 I | 20 | -0.564-14.9 I |
| 6 | -0.049-14.7 I | 21 | 0.544-14.8 I |
| 7 | 0.062-9.04 I | 22 | 33.2-31.4 I |
| 8 | 0.005-17.4 I | 23 | -0.291-14.5 I |
| 9 | 0.051-14.1 I | 24 | 0.183-15.1 I |
| 10 | -0.258-6.58 I | 25 | 1.72-50.0 I |
| 11 | -0.226-16.4 I | 26 | 0.139-14.4 I |
| 12 | 0.239-14.0 I | 27 | -0.446-15.3 I |
| 13 | 0.224+0.69 I | 28 | -16.52-26.7 I |
| 14 | -0.465-15.9 I | 29 | 0.654-14.4 I |
| 15 | 0.468-14.1 I | 30 | -2.96-16.0 I |

(a)  (b)

**FIGURE 3**

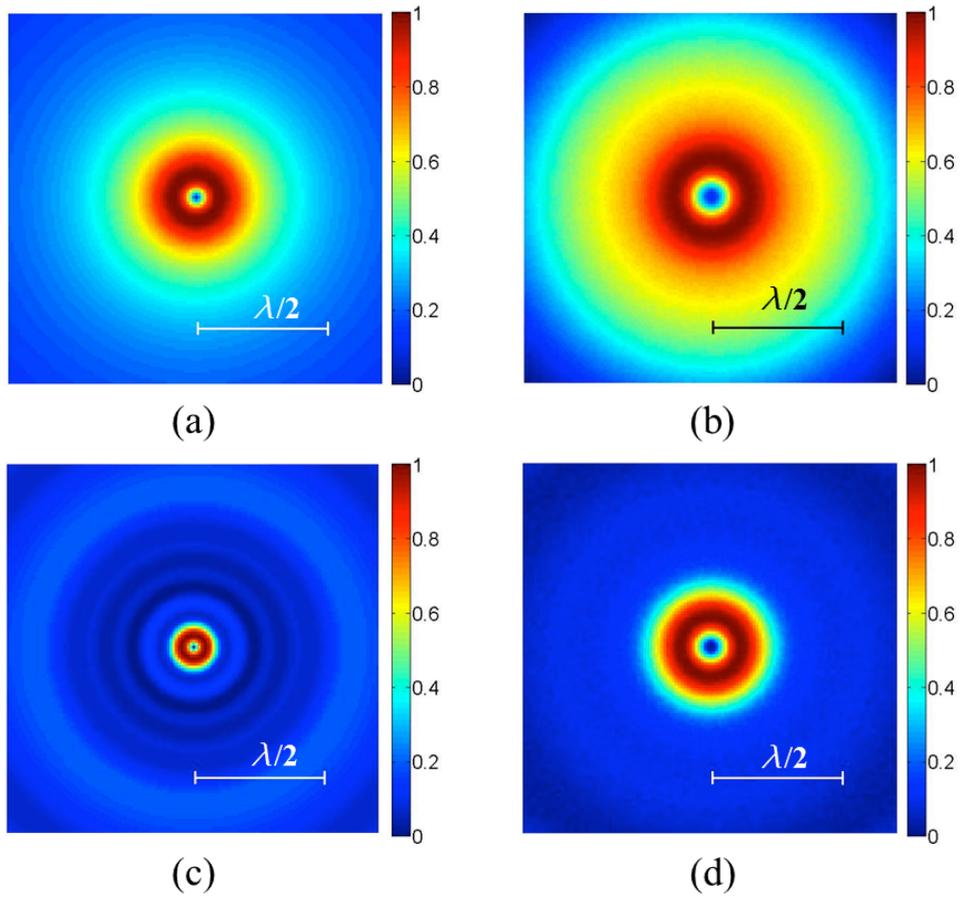

(a) (b) (c) (d)

**FIGURE 4**

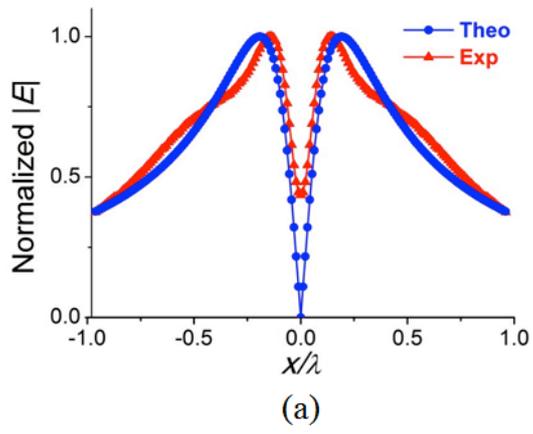 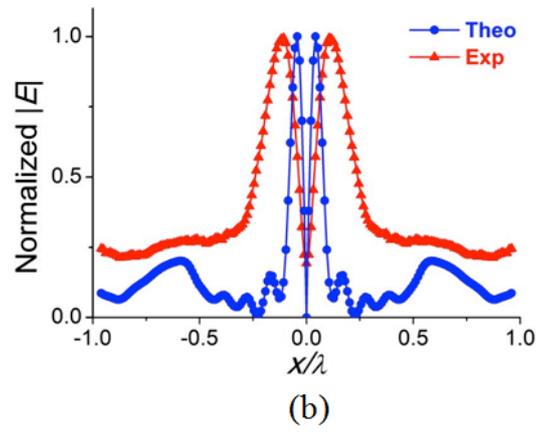

(a) (b)

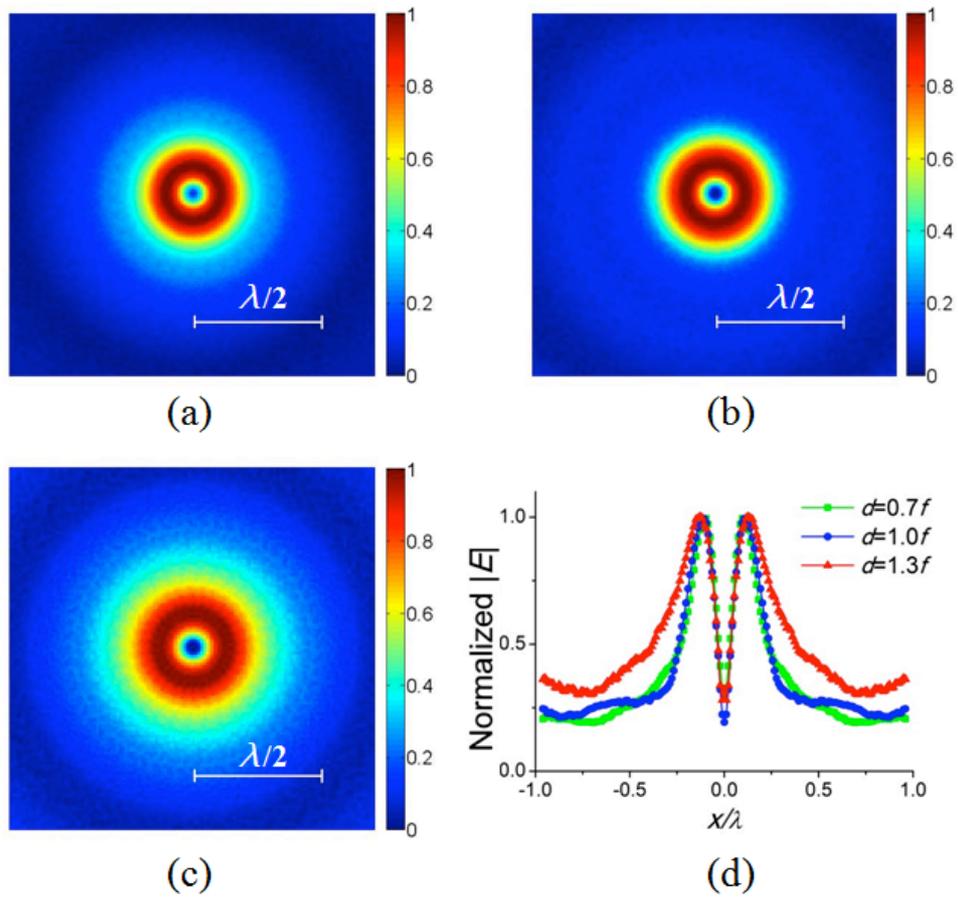